\documentclass[aps,preprint]{revtex4}

\begin{document}

\bibliographystyle{prsty}

\draft

%\wideabs{

\title{ Comment on "Magnus force and acoustic Stewart-Tolman effect in
type-II superconductors, by Fil {\it et al.}" }

\author{ P. Ao }
\address{ Departments of Mechanical Engineering and Physics,
          University of Washington, Seattle, WA 98195, USA }
%\address{ }
\date{ June 18, 2007 }
%\date{\today}

%\maketitle

\begin{abstract}
 Fil {\it et al.} has proposed an interesting experimental method to
 investigate vortex dynamics. Some preliminary results have been
 obtained. In this comment I discuss a few missing but strongly related
 theoretical models and experiments on Hall anomaly and
 Magnus force. I conclude that those missing literature can enhance
 the value of novel experimental method proposed in the commented
 2006 Europhysics Letters by Fil {\it et al.}.  \\
 {\ }  
\end{abstract}

%\pacs{PACS numbers: 74.25.Op }
% Mixed states, critical fields, and surface sheaths.

%}

\maketitle

In their letter Fil et al  \cite{fil06}  presented a novel
experimental method to check various dynamical elements in the
fundamental equation of vortex. It is likely that new information
on vortex dynamics can be obtained, which may help to resolve
outstanding puzzles. Nevertheless, a few key facts were presented
wrong, which will make the interpretation of their data difficult,
if not introducing additional confusing. Here I wish to point out
such mistakes.

1) A well-known microscopic theory of effective Magnus force on a
single moving vortex predicts its magnitude is greatly reduced by
the relaxation time due to impurity at length scale smaller than
the coherence length, which does not agree with the expectation of
classical limit. This theory was summarized in Ref. [2] of Fil et
al..

However, it was pointed out that the use of relaxation time
approximation is wrong in this context \cite{ao98}. In additional,
a competing microscopic theory has reached the opposite conclusion
that the magnitude of the Magnus force should be what expected in
the classical limit in the light of two fluid type model
\cite{at03}, and both the Magnus force and the vortex friction
have been derived without the relaxation time approximation
\cite{at03,tan96}, further extended by others \cite{ns06}. Even if
Fil et al believe such a competing theory should be wrong, it
should be cited and discussed, and would even be better to be
proved wrong if their data could do it.

2) Among various experiments related to Magnus force in
superconductors, two types should be particularly relevant here.
One was a direct measurement of the Magnus force on moving
vortices. The data indicated an agreement with the competing
theory \cite{zhu97}. Another was to check the effective Magnus
force in situations there is no change in relaxation time, but
other factors are varying \cite{ghenim04}. The data indicated a
disagreement with the relaxation time controlled Magnus force
theory. However, those two very relevant experiments were not
cited by Fil et al.

3) One of most difficult problems has been the explanation of the
Hall anomaly in superconductors: With the Magnus force as big as
what expected in the classical limit, how could the Hall angle is
usually not only small, but often changes its sign?  One solution
to this puzzle was proposed by myself in 1995 \cite{ao95}, to
consider the vortex many-body effect, combining with the pinning
effect on the scale larger than coherence length. Such idea was
rediscovered by Kopnin and Vinokur in 1999 \cite{kv99} (Ref.[3] in
Fil et al.). However, because neither Kopnin and/or Vinokur nor
Fil et al discussed those prior references, it is wrong to credit
only to Kopnin and Vinokur for such an idea, when Fil et al stated
"According to the current theoretical conceptions, this effect may
be … of macroscopic one, caused by a transverse force that may
emerge at large (much larger than the core size) displacements of
the vortices in the pinning potential" with reference to Kopnin
and Vinokur.

Incidentally, it should be mentioned that Kopnin and Vinokur
\cite{kv99} did credit the experimental measurement \cite{zhu97}
of the Magnus force.

4) There are various types of vortex-phonon interactions. One was
indicated in a quantitative way in 1994 \cite{nat94}, existing
even in zero temperature limit. It has been elaborated in various
experimental checkable situations \cite{dan96}.  It is my
understanding that the major goal of Sonin \cite{sonin97} (Ref.[1]
in Fil et al.) was to disprove the existence of such vortex-phonon
effect. In the light vortex-phonon interaction method proposed by
Fil et al., it does not appear appropriate to only cite Sonin.

In conclusion, without citation and discussion of those missing
theories and experiments, the proper interpretation of Fil et al
data will be difficult. It is not appropriate for Fil et al. to do
that if they had known those works. If Fil et al were not aware of
those works, the inclusion of them should enhance the value of
their experimental effort.

\end{document}